\documentclass[aps,prd,nofootinbib,10pt,tightenlines,twocolumn,noti,floatfix,superscriptaddress,showkeys]{revtex4-2}

\usepackage{lipsum}

\usepackage{float}

\usepackage{subcaption} 
\usepackage{caption}

\usepackage[utf8]{inputenc}          
\usepackage{graphicx}         
\usepackage[usenames,dvipsnames]{xcolor}

\usepackage{comment}
\usepackage{amsmath,amssymb,amsfonts,slashed} 
\usepackage{mathtools}          
\usepackage[linktocpage,breaklinks]{hyperref}
\usepackage{mathrsfs}
\usepackage{txfonts}
\usepackage{bm}
\usepackage{stmaryrd}
\usepackage{tensor}
\usepackage[utf8]{inputenc}

\usepackage{orcidlink}
\usepackage{epsfig}
\usepackage{epstopdf}

\usepackage{placeins}

\usepackage{cleveref}

\definecolor{mred}{RGB}{127,0,25}
\definecolor{mdgr}{RGB}{51,51,51}
\definecolor{mag}{RGB}{211, 54, 130}
\definecolor{verm}{RGB}{164, 25, 0}

\hypersetup{colorlinks=true,
            citecolor=NavyBlue,
            linkcolor=NavyBlue,
            urlcolor=NavyBlue}
\usepackage{siunitx}                                  
\DeclareSIUnit{\fm}{\femto\metre}                     

\usepackage{comment}

\newcount\hh
\newcount\mm
\mm=\time
\hh=\time
\divide\hh by 60
\divide\mm by 60
\multiply\mm by 60
\mm=-\mm
\advance\mm by \time
\def\hhmm{\number\hh:\ifnum\mm<10{}0\fi\number\mm}

\def\be{\begin{equation}}
\def\ee{\end{equation}}

\def\tE{\tilde{E}}
\def\tL{\tilde{L}}
\def\tr{\tilde{r}}

\begin{document}

\preprint{APS/123-QED}

\title{Beyond the Separatrix: Analytic Continuation of Darwin Variables\\ for Plunging Geodesics in Schwarzschild Spacetime}

\author{Francisco M. Blanco\orcidlink{0000-0002-7711-8395}}
\affiliation{Max Planck Institute for Gravitational Physics (Albert Einstein Institute), Am M\"{u}hlenberg 1, Potsdam, 14476, Germany}


\begin{abstract}

We study geodesic motion of a test particle in Schwarzschild spacetime. Bound and scattering geodesics are commonly described using Darwin variables, which provide a convenient parametrization of the radial motion. However, this description breaks down at the separatrix and does not extend straightforwardly to plunging trajectories. We construct an analytic continuation of Darwin variables that yields a real parametrization of bound, scattering, and plunging Schwarzschild geodesics, thereby providing a unified kinematical description of all types of test-mass motion. As a proof of concept, we then apply these variables to a simple non-geodesic evolution in which the energy and angular momentum are driven by a constant external force. This toy model is not intended to represent a physical radiation-reaction model, but rather to illustrate how the extended variables can be used to follow an orbit through a transition to plunge using a single orbital phase variable across the separatrix.

\end{abstract}

\maketitle

\vspace{0.1cm}

\section{Introduction}

The detection of gravitational waves by the LIGO and Virgo collaborations \cite{introLIGO1,introLIGO2,introLIGO3} has opened a new observational window onto the strong-field regime of general relativity. Future detectors such as LISA \cite{introLISA1} will extend this reach, making it possible to observe long-lived extreme mass-ratio inspirals (EMRIs), in which a stellar-mass object spirals into a supermassive black hole. These systems probe the geometry of spacetime deep within the strong-field region and provide some of the most precise tests of general relativity that will be available in the coming decades \cite{pound}.

The gravitational waveform from a binary system typically proceeds through four phases: inspiral, plunge, merger, and ringdown. In the context of extreme mass-ratio inspirals, the system evolves adiabatically for most of its inspiral phase \cite{flanagan2}. Hence, it is possible to describe the trajectory of the secondary as a geodesic with slowly varying orbital parameters. For Schwarzschild and Kerr spacetimes, the Darwin (or “Keplerian”) coordinates $(\chi,p,\epsilon)$ have proven particularly convenient for describing this regime: they parameterize bound motion in terms of orbital elements and a radial phase variable that evolves monotonically with proper time \cite{darwin}. They can also describe scattering trajectories.

However, as the inspiral approaches the separatrix —the boundary between bound orbits and plunging trajectories— this description breaks down. Near this transition, the radial period diverges, the resulting dynamics are no longer well-approximated by adiabatic evolution, and the two-timescale expansion must be supplemented by a description of the transition regime between the \emph{last stable orbit} and the onset of plunge. Early insight into the transition to plunge of extreme mass-ratio inspirals was provided by the classical Ori–Thorne model \cite{PhysRevD.62.124022}, which describes the transition for circular inspirals by matching the orbital phase to a local expansion of the effective radial potential near the ISCO. Similar work was done in the comparable mass regime on \cite{PhysRevD.62.064015}.

Within the gravitational self-force framework, several complementary approaches have recently been developed to characterize this transition. Lhost and Compère \cite{Lhost:2024jmw} derived analytic late-time expressions for eccentric inspirals near the separatrix in terms of self-force quantities evaluated there. Becker and Hughes \cite{Becker.PhysRevD.111.064003} constructed a smooth continuation from the last stable bound geodesic into plunge, highlighting the weakening of the radial restoring force. Küchler et al. \cite{Kuchler:2024esj,Honet:2025dho,Küchler_2026} exploited the hierarchy of orbital, transition, and radiation-reaction timescales to build a controlled multiscale expansion incorporating the early plunge for quasicircular orbits. Finally, Faggioli et al. \cite{Faggioli.xq3j-4jtx} studied the waveform imprint of critical plunges in Kerr, showing how the dominant modes are organized across spin and eccentricity.

In this work, we take a complementary viewpoint. Rather than modeling the dynamics of the transition, we focus on the kinematics of the underlying geodesic motion and seek a parametrization that remains well defined across all regimes. Standard Darwin variables fail when the radial potential ceases to admit two real turning points. We show that this obstruction can be overcome by extending the variables $(p,\epsilon,\chi)$ into the complex plane while keeping the physical quantities $(r,E,L)$ real. Guided by this observation, we construct an \emph{extended Darwin map} that provides a real parametrization of bound, scattering, and plunging geodesics within a common framework. The construction exploits the multivalued relation between $(p,\epsilon)$ and $(E,L)$, allowing the different geodesic sectors to be described continuously within the same set of variables.

When a weak driving force pushes the system into plunge, the parametrization develops a nonanalytic behavior at the separatrix. This appears as a square-root kink associated with the merger of the radial turning points. The kink reflects a coordinate artifact rather than a physical singularity. To remove it, we regularize one of the radial roots with a smoothing function involving a length scale $l$, which controls only the sharpness of the coordinate transition near the separatrix. For generic crossings, the resulting motion is insensitive to the precise value of $l$. Dependence on $l$ arises only in the finely tuned case where the crossing occurs parametrically close to periapsis. Away from this nongeneric regime, the parametrization remains smooth and independent of the arbitrary parameter $l$.

The extended Darwin variables provide a natural description of motion through both inspiral and plunge, and offer a convenient setting for incorporating radiation-reaction effects through fluxes on $(p,\epsilon)$. They may also be useful for future effective-one-body constructions \cite{Buonanno_1999,PhysRevD.62.064015,Damour:2000we,Buonanno:2025xu} by supplying coordinates that remain regular through the separatrix.

The paper is organized as follows. In Section \ref{sec:schwarzschildphase}, we review the equations of motion of a test particle in the Schwarzschild metric and study the phase space of geodesic motion using energy and angular momentum. In Section \ref{sec:darwin}, we review the use of Darwin variables and their relation to energy and angular momentum. We construct the phase space in these variables and explain their domain of validity. Specifically, we show how semi-latus rectum and eccentricity become complex below the separatrix. In Section \ref{sec:analyticcont}, we extend Darwin variables by analytically continuing the relativistic anomaly $\chi$. We show that making all Darwin variables complex provides enough freedom to keep the dynamics real. In Section \ref{sec:extendeddarwin}, we define the extended Darwin map and show how it can be used to describe all types of geodesic motion. In Section \ref{sec:rad}, we show how to construct the extended variables in a way that transitions smoothly from bound/scattering trajectories to plunges in the presence of a small driving force. We show how the parametrization fails to be smooth if the separatrix is crossed too close to the periapsis passage. Finally, we model the effect of a constant driving force by evolving energy and angular momentum at a fixed rate. We then describe the system as evolving through a series of geodesics with changing orbital parameters. We emphasize that the construction provides a generalized orbital phase variable that remains continuous across the separatrix. This variable extends the role played by the relativistic anomaly for bound motion and may offer a way to track the evolution of the waveform phase through the transition from bound motion to plunge.

\section{The Phase space of Schwarzschild geodesics}\label{sec:schwarzschildphase}

Consider a point particle of mass $m$ moving along a geodesic of the Schwarzschild metric with spacetime interval

\begin{equation}
    ds^2 = -(1-\frac{2M}{r})dt^2 +(1-\frac{2M}{r})^{-1}dr^2 +r^2 d\Omega^2.
\end{equation}
Here, $M$ is the mass of the central black hole, $d\Omega^2 = \sin^2\theta d\varphi^2 +d\theta^2$ and we have set $G=c=1$.

Using the symmetries of the Schwarzschild metric, we can set $\theta=\pi/2$ and find equations for the remaining variables in terms of the conserved energy and angular momentum. We use a tilde to define conserved quantities per unit mass so that $\tilde{r}=r/M$, $\tilde{\tau}=\tau/M$, $\tilde{E}=E/m$ and $\tilde{L}=L/(mM)$. The equations of motion are
\begin{subequations}
    \begin{eqnarray}
        \frac{d\tilde{t}}{d\tilde{\tau}}&=&\frac{\tilde{E}}{1-\frac{2}{\tilde{r}}},\\
        \frac{d\varphi}{d\tilde{\tau}} &=& \frac{\tilde{L}}{\tilde{r}^2},\\
        \frac{d\tilde{r}}{d\tilde{\tau}}&=&\pm \sqrt{R(\tilde{r};\tilde{E},\tilde{L})}\label{eq:radialel}.
    \end{eqnarray}
\end{subequations}
Here, we defined the radial function
\begin{equation}
\begin{aligned}
    R(\tilde{r};\tilde{E},\tilde{L})&=\tilde{E}^2 -\big(1-\frac{2}{\tr}\big)\big(1+\frac{\tilde{L}^2}{\tilde{r}^2}\big)\\
    &=\tilde{E}^2 - V_{\text{eff}}(\tilde{r};\tilde{L})
\end{aligned}
\end{equation}
The roots of $R(\tilde{r})$ are the turning points of radial motion, where $\tilde{E}^2$ intersects $V_{\text{eff}}$ and $d\tilde{r}/d\tilde{\tau}=0$. For $\tilde{E}^2<1$, $R$ can have three positive real roots
\begin{equation}
    \tilde{r}_1<\tilde{r}_2<\tilde{r}_3.
\end{equation}
If all three roots exist, motion can occur in two ways. The particle's radius can oscillate between $[\tilde{r}_2,\tilde{r}_3]$, which corresponds to eccentric bound orbits between periapsis and apoapsis. The particle can also be confined to the interval $[0,\tilde{r}_1]$, which corresponds to plunging motion with a turning point at $\tilde{r}_1$. Changing values of energy $\tE$ and angular momentum $\tL$ will produce one of three transitions. First, $\tilde{r}_1$ and $\tilde{r}_2$ can merge at the unstable circular orbit (UCO) and then become complex conjugate roots. Second, $\tilde{r}_2$ and $\tilde{r}_3$ can merge at the stable circular orbit (SCO) and then become complex conjugate roots. Third, all three roots merge at the innermost stable circular orbit (ISCO), leaving two complex conjugate roots and a real root, which becomes the turning point for plunging motion. The roots of $R(\tilde{r};\tilde{E},\tilde{L})$ are shown in Fig. \ref{fig:five_subfigs} for different values of $\tilde{E}^2$ and $\tilde{L}^2$. Note that these transitions might occur multiple times, depending on the physical evolution of $\tE$ and $\tL$ which can conspire to make the system go back and forth between different phases.

If only one real root exists, then geodesics are always plunges. If there are only two real roots $\tr_1$ and $\tr_2$, the particle can move in two ways. It can move on a scattering trajectory, starting at infinity, reaching $\tr_2$ and then leaving to infinity again or it can also move between $[0,\tr_1]$. In that case, the particle plunges from $\tr_1$ into the central black hole. We note that the effective potential $V_{\text{eff}}$ has a potential barrier for all $\tilde{L}^2>12$, which is necessary for having three real roots. 

Circular orbits are determined by the twofold requirement $R=\partial_{\tilde{r}} R=0$. This determines a curve in the parameter space $(\tE,\tL)$, defined by $\tilde{E}^2 = V_\text{min}(\tilde{L})$. Here, $V_\text{min}(\tL)$ is the value of the potential $V$ at the SCO radius for a given value of angular momentum. Similarly, the curve defined by $\tilde{E}^2 = V_{\text{max}}(\tilde{L})$ describes unstable circular orbits, situated at the top of the potential barrier. Here, $V_\text{max}$ is the value of the potential $V$ at the UCO. This curve is also known as the \emph{separatrix} and it separates bound/scattering trajectories from plunges.

\begin{figure}[h]
    \centering
\includegraphics[width=1\linewidth]{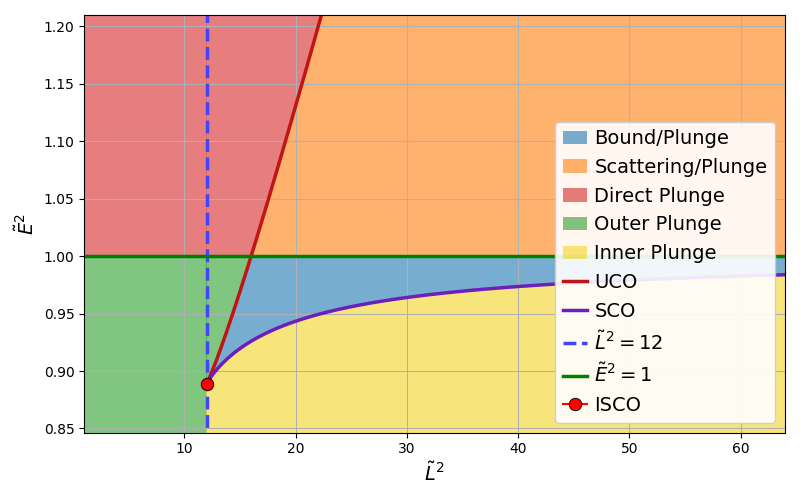}
    \caption{Phase space of Schwarzschild geodesics in $(\tilde{E}^2,\tilde{L}^2)$ variables. }
    \label{fig:ELphasespace}
\end{figure}

\begin{figure*}[t]
\centering
  \begin{subfigure}[t]{0.3\textwidth}
    \centering
    \includegraphics[width=\linewidth]{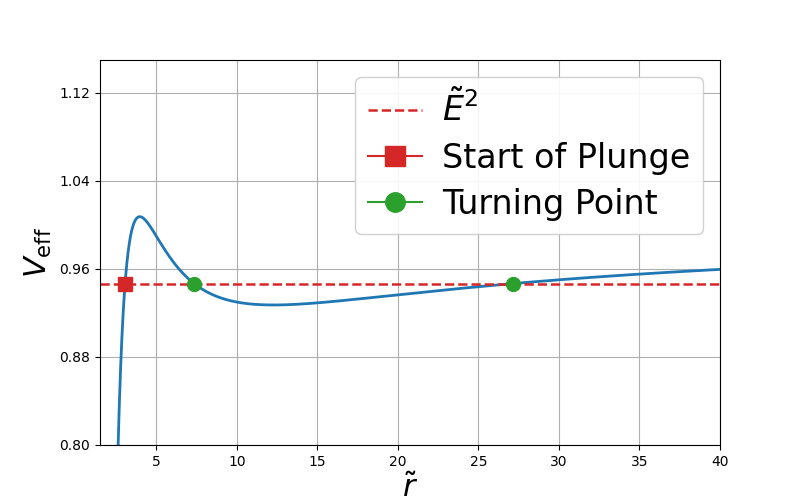}
    \caption{Bound/Plunge Phase.}\label{fig:sub1}
  \end{subfigure}\hfill
  \begin{subfigure}[t]{0.3\textwidth}
    \centering
    \includegraphics[width=\linewidth]{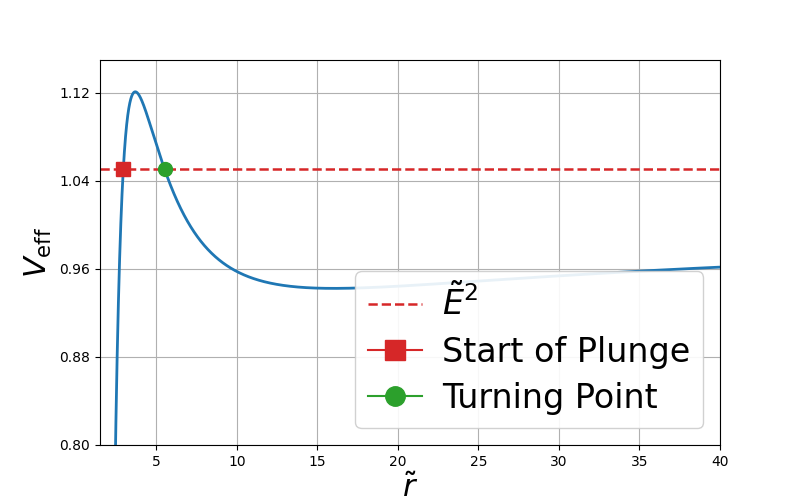}
    \caption{Scattering/Plunge Phase.}\label{fig:sub2}
  \end{subfigure}\hfill
  \begin{subfigure}[t]{0.3\textwidth}
    \centering
    \includegraphics[width=\linewidth]{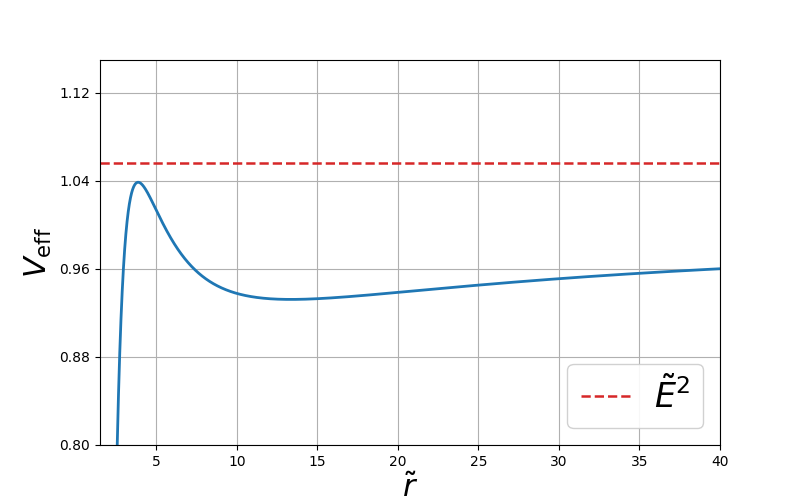}
    \caption{Direct Plunge Phase.}\label{fig:sub3}
  \end{subfigure}

  \vspace{0.5em}

  \begin{subfigure}[t]{0.3\textwidth}
    \centering
    \includegraphics[width=\linewidth]{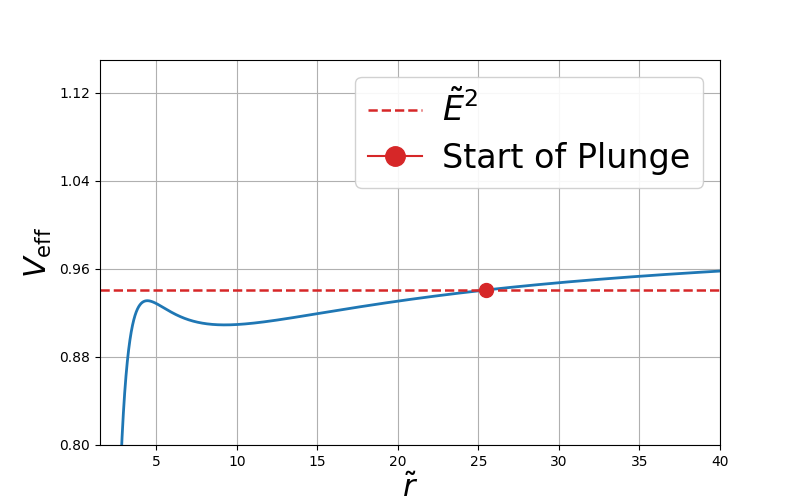}
    \caption{Outer Plunge Phase.}\label{fig:sub4}
  \end{subfigure}\hspace{0.08\textwidth}
  \begin{subfigure}[t]{0.3\textwidth}
    \centering
    \includegraphics[width=\linewidth]{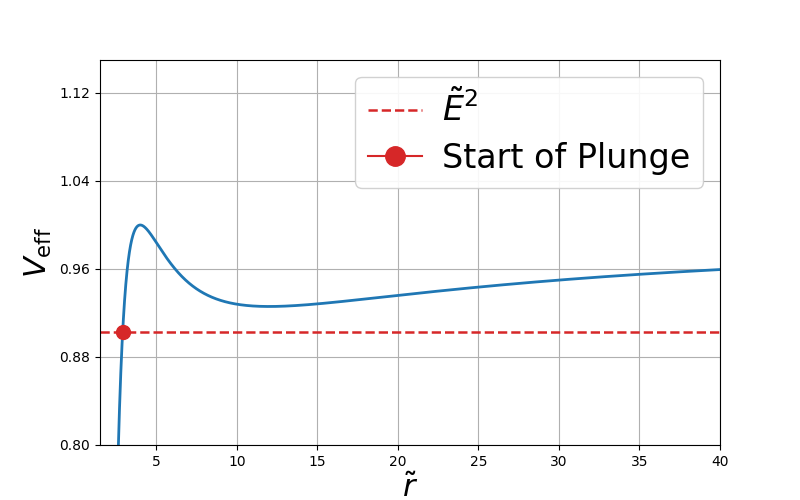}
    \caption{Inner Plunge Phase.}\label{fig:sub6}
  \end{subfigure}

  \caption{Effective potential $V_{\text{eff}}(\tilde r;\tilde L)$ and energy squared $\tilde E^2$ for representative points in the different phase space regions in Fig.~\ref{fig:ELphasespace}. Turning points are depicted in green, while plunge starting points are depicted in red. (a) There are two types of geodesics: \textbf{bound orbits} and \textbf{inner plunges}. The two families of orbits become connected at separatrix. (b) There are two types of geodesics: \textbf{Scattering trajectories} and \textbf{inner plunges}. The two families of trajectories become connected at separatrix. (c) All geodesics are \textbf{direct plunges} from infinity. (d) All geodesics are \textbf{outer plunges} from a finite turning point. (e) All geodesics are \textbf{inner plunges}. As opposed to the inner plunges in the Bound/Plunge phase, these inner plunges are disconnected from bound orbits.}
  \label{fig:five_subfigs}
\end{figure*}

Geodesic motion can be divided into different ``phases" depending on the number of roots of $R(\tilde{r})$. In Fig. \ref{fig:ELphasespace}, we show all possible phases. A typical effective potential showing possible orbits and turning points for each region can be seen in Fig. \ref{fig:five_subfigs}. We classify trajectories as bound, scattering, or plunging. Bound trajectories oscillate between $\tr_2$ and $\tr_3$, while scattering trajectories extend from $\tr_2$ to infinity. Plunging trajectories are further classified as inner, outer, or direct according to whether their maximum radius is $\tr_1$, $\tr_3$, or infinity, respectively.

\begin{itemize}
    \item The \textbf{Bound/Plunge} phase corresponds to  $V_\text{min}(\tilde{L})<\tilde{E}^2<V_\text{max}(\tilde{L})$ with $\tilde{E}^2<1$. There are two distinct types of geodesics in this region: Bound orbits that oscillate between the two greater roots $\tilde{r}_2$ and $\tilde{r}_3$, and  inner plunges that start on the smallest root $\tilde{r}_1$, to the left of the potential barrier.
    \item The \textbf{Scattering/Plunge} phase corresponds to $1<\tE^2<V_\text{max}(\tL)$. There are also two distinct types of geodesics in this region: scattering trajectories that start at infinity and bounce off at the root $\tilde{r}_2$, and inner plunges that start on the smaller root $\tilde{r}_1$, left of the potential barrier. 
    \item The \textbf{Direct Plunge} phase corresponds to $1<\tE^2$ (and $V_\text{max}(\tL)<\tE^2$ if a potential barrier exists). All geodesics in this region are direct plunges that start from infinity. Above the dashed blue line, there exists a potential barrier, but this doesn't affect the motion since energy is greater than it.
    \item The \textbf{Outer Plunge} phase corresponds to $V_{\max}(\tL)<\tE^2<1$ whenever a potential barrier exists, or just $\tE^2<1$ when it does not. All geodesics in this region are outer plunges that start at the greater root $\tilde{r}_3$. Similarly to the previous region, above the dashed blue line, there exists a potential barrier, but the energy is always greater than it so it doesn't affect what types of orbits we see.
    \item The \textbf{Inner Plunge} phase corresponds to energies less than the minimum of the potential, which means that the particle is plunging from the left side of the potential barrier. All geodesics are inner plunges.
\end{itemize}

\section{Darwin Variables} \label{sec:darwin}

In this section, we review the application of Darwin variables to geodesic motion in Schwarzschild spacetimes. The direct evaluation of integrals in the variables $(r,E,L)$ can present complications since $r$ is an oscillating variable whose rate of change vanishes at each turning point. It is therefore convenient to find new coordinates where the evolution becomes monotonic. 

The bound orbits in Fig. \ref{fig:sub1} and scattering trajectories in Fig. \ref{fig:sub2} can be described in such coordinates using \emph{Darwin variables} $(\chi,p,\epsilon)$. These are defined by
\begin{equation}\label{eq:darwinparam}
    \tr=\frac{p}{1+\epsilon \cos \chi}.
\end{equation}
Here, $\chi$ is the relativistic anomaly\footnote{The name comes from the fact that $\chi$ differs from the azimuthal angle $\varphi$ in General Relativity, thus representing a measurable ``anomaly" that can distinguish Newtonian and relativistic dynamics.}, $p$ is the semi-latus rectum, and $\epsilon$ is the eccentricity. These coordinates cover only part of phase space: for \(\epsilon<1\) they describe bound orbits, while for \(\epsilon>1\) they describe scattering trajectories. In other words, they parameterize one of the two possible types of geodesics on the bound and scattering regions in Fig. \ref{fig:ELphasespace}.

For $\epsilon<1$, the radius lies between periapsis and apoapsis:
\begin{subequations}\label{eq:periapo}
\begin{eqnarray}
    \tr_+&=&\frac{p}{1+\epsilon},\\
    \tr_-&=&\frac{p}{1-\epsilon}.
\end{eqnarray}
\end{subequations}
For $\epsilon>1$, $r_-$ becomes negative, indicating that the particle travels from infinity to $\tr_+$ and then back to infinity. Even though $\tr_\pm$ coincide with the roots defined earlier as $\tr_2$ and $\tr_3$, we reserve the $\pm$ notation for roots written in terms of $(p,\epsilon)$ for reasons that will become clear in Section \ref{subsec:turningpoints}.

This parametrization is advantageous in that $\chi$ is a monotonically increasing function of $\tau$
\begin{equation}\label{eq:chitopropertime}
    \frac{d\chi}{d\tilde{\tau}} = \frac{\pm \sqrt{R}}{d\tr/d\chi}.
\end{equation}
The $\pm$ in the numerator has to be chosen by hand at each branch of Eq. (\ref{eq:radialel}). This exactly cancels with the sign changes in $d\tr/d\chi$. Furthermore, both the numerator and denominator vanish at turning points, making $d\chi/d\tilde{\tau}$ finite and monotonic throughout the orbit. This is the defining feature of a good \emph{phase} variable, namely, a parameter that increases continuously along the trajectory and can therefore be used as an orbital phase variable for the emitted waveform.

It is important to note that there are multiple alternatives to Eq. (\ref{eq:darwinparam}). As long as the phase variable is monotonic and produces the correct radial turning points (cf. Eq. (\ref{eq:radialel})), any parametrization is valid. In other words, other than the turning points, there is no dynamical information in Eq.(\ref{eq:darwinparam}): It is a change of coordinates, not a solution to the equations of motion. The dynamics are incorporated once we return to proper time by using Eq. (\ref{eq:chitopropertime}), where the radial function $R$ determines the flow of the phase $\chi$ with proper time $\tilde{\tau}$.

\subsection{Map between $(\tE,\tL)$ and  $(p,\epsilon)$}

The relation between $(\tE,\tL)$ and $(p,\epsilon)$ follows by substituting Eqs.~\eqref{eq:periapo} into Eq.~\eqref{eq:radialel} and using that the radial velocity vanishes at turning points. The mapping is
\begin{subequations}\label{eq:eltopepsilon}
\begin{eqnarray}
    \tE^2&=&\frac{(p-2)^2-4\epsilon^2}{p\big[p-3-\epsilon^2\big]}\label{eq:petoE},\\
    \tL^2&=&\frac{p^2}{p-3-\epsilon^2}.\label{eq:petoL}
\end{eqnarray}
\end{subequations}
In what follows, we analyze which values of $(p,\epsilon)$ correspond to which values of $(\tE,\tL)$.

As the Darwin variables approach \(p=3+\epsilon^2\), both \(\tE^2\) and \(\tL^2\) approach infinity; and then become negative, which signals an unphysical region. The curve $p=3+\epsilon^2$ corresponds to the null limit of timelike geodesics, which is why energy and angular momentum diverge there. 
Solving for $\tE^2=1$, we obtain the condition 
\begin{equation}\label{eq:e=1cond}
   (p-4)(\epsilon^2-1)=0
\end{equation}
which determines the boundaries between bound and scattering motion. The curve $p=4$ is not in the typical domain of the Darwin variables, but we add it here for completeness.

Solving for $\tL^2=12$, which determines when the potential barrier drops, we find the condition
\begin{equation}\label{eq:l2=12cond}
    p=6\pm i \sqrt{12}\epsilon.
\end{equation}
Therefore, real Darwin coordinates always parametrize geodesics with angular momentum greater than $\tL^2=12$, except at the ISCO, where $p=6$ and $\epsilon=0$. 

The Jacobian of the coordinate transformation $(\tE,\tL)\rightarrow (p,\epsilon)$ is
\begin{equation}\label{eq:jacobian}
    \mathcal{J}=\frac{2\epsilon\big[p-6-2\epsilon\big]\big[p-6+2\epsilon\big]}{\big[p-3-\epsilon^2\big]^3}.
\end{equation}
We see that the Jacobian becomes singular at $\epsilon=0$ and at $p=6\pm2\epsilon$. The singularity of the Jacobian reflects a degeneracy of the map: different curves in $(p,\epsilon)$ space correspond to the same values of $(\tE,\tL)$.

The line $\epsilon=0$ corresponds to circular orbits: for $p>6$ these are stable, while for $3<p<6$ they are unstable. These satisfy $\tE^{2}=V_{\min}(\tL)$ and $\tE^{2}=V_{\max}(\tL)$, respectively. 
The curve $p=6+2\epsilon$ corresponds to the LSO\footnote{The last stable orbit is a bound orbit whose periapsis coincides with the maximum of the potential barrier. In the study of dynamical systems, these are also known as homoclinic curves \cite{Levin:2008yp,Grossman:2008yk}.}. The LSO has the same energy and angular momentum as an unstable circular orbit (i.e., $\epsilon=0$ and $3<p<6$). In other words, the LSO curve and the unstable circular orbit curve are distinct in $(p,\epsilon)$ space, but they coincide in energy and angular momentum. The curve $p=6+2\epsilon$ is also known as the ``separatrix" since it splits the dynamics between bound/scattering trajectories and plunges. 
The curve $p=6-2\epsilon$ is outside the typical domain of validity of the Darwin variables, but it corresponds to the energy and angular momentum of a stable circular orbit (i.e., $\epsilon=0$ and $6<p$).

\subsection{Branches for $(p,\epsilon)$}\label{sec:branches}

The mapping in Eq. (\ref{eq:eltopepsilon}) can be inverted to obtain $p$ and $\epsilon$ in terms of $\tE^2$ and $\tL^2$. First, we solve Eq. (\ref{eq:petoL})
\begin{equation}\label{eq:epssolcub}
    \epsilon^2 =p-\frac{p^2}{\tL^2}-3.
\end{equation}
We can insert this equation into Eq. (\ref{eq:petoE}) to obtain a cubic polynomial for $p$
\begin{equation} \label{eq:pcubic}
    \frac{\tE^2}{\tL^2}p^3-(1+\frac{4}{\tL^2})p^2+8p-16=0.
\end{equation}
Each of the three solutions to Eq. (\ref{eq:pcubic}) can be inserted into Eq. (\ref{eq:epssolcub}) to obtain the three branches of $p$ and $\epsilon$. In other words, for any given value of $(\tE,\tL)$ there are three possible branches for $(p,\epsilon)$.

If a polynomial with real coefficients has complex roots, they must come in conjugate pairs. Therefore, either all three branches of $p$ are real, or one is real and the other two are complex conjugates. Note that $\epsilon$ is derived directly from Eq. (\ref{eq:epssolcub}) and not from a cubic polynomial, so all three branches of $\epsilon$ can be complex. Cubic polynomials have analytic solutions in terms of Cardano's formula, although their details are not important here. We note that the discriminant of the cubic polynomial $D$ determines whether there exist real or complex branches. When $D<0$, there exist three distinct real branches for $p$. When $D=0$, there exist three real roots with at least two equal (double or triple root). When $D>0$, the branches that merged become complex conjugates. $D=0$ corresponds exactly with the separatrix and marks the transition to plunge. 

If we focus on energies and angular momenta corresponding to bound/scattering trajectories, we see that the three branches of $(p,\epsilon)$ are organized as follows: The \emph{first branch} satisfies $p>6+2\epsilon$, which is the typical domain of the Darwin variables. The \emph{second branch} $p-2\epsilon<p<6+2\epsilon$ and the \emph{third branch} of $(p,\epsilon)$ satisfies $p<6-2\epsilon$.

\begin{figure}[t]
\centering
\begin{subfigure}[t]{0.42\textwidth}
    \centering
    \includegraphics[width=\linewidth]{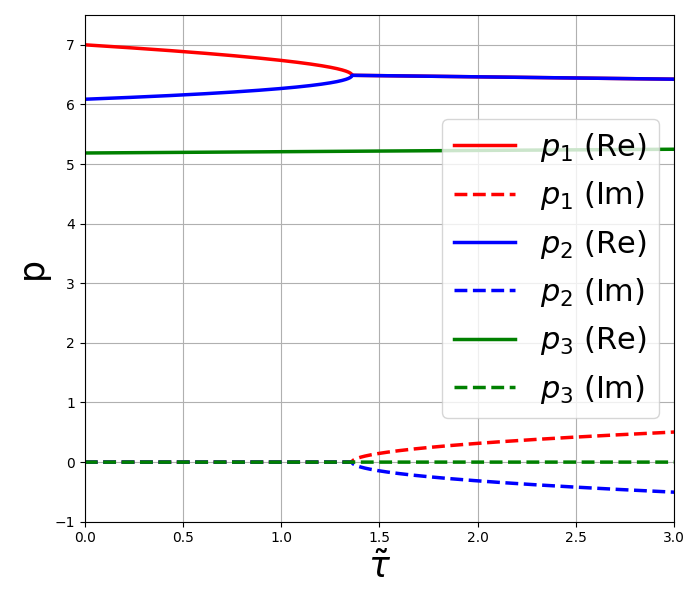}
    \caption{Evolution of each branch of semi-latus rectum $p$.}
    \label{fig:pbranches}
\end{subfigure}\hfill
\begin{subfigure}[t]{0.42\textwidth}
    \centering
    \includegraphics[width=\linewidth]{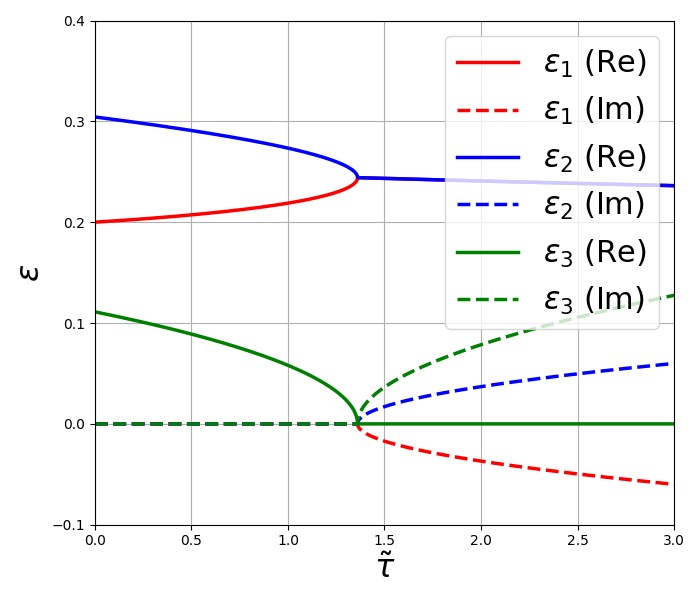}
    \caption{Evolution of each branch of eccentricity $\epsilon$.}
    \label{fig:eccbranches}
\end{subfigure}

\caption{Evolution of all three branches of $(p,\epsilon)$ obtained by varying energy and angular momentum across the separatrix.}
\label{fig:pebranches}

\end{figure}

In Fig. \ref{fig:pebranches}, we vary $(\tE,\tL)$ from across the separatrix and show the evolution of all three branches of $(p,\epsilon)$. The first, second and third branches are depicted in red, blue and green, respectively. As energy and angular momentum transition from the bound/scattering phases into a plunge, the \emph{first} and \emph{second} branches of $p$ become complex conjugates, while the \emph{third} branch remains real. All branches of $\epsilon$ become complex after the separatrix is crossed. We use constant fluxes of $\tE$ and $\tL$ in all figures.

\subsection{Turning points}\label{subsec:turningpoints}

By inserting Eqs. (\ref{eq:darwinparam}) and (\ref{eq:eltopepsilon}) into Eq. (\ref{eq:radialel}), it is easy to check that the radial function $R(\tr;\tE,\tL)$, written in Darwin variables, possesses three roots.
\begin{subequations}\label{eq:rootsinpe}
    \begin{eqnarray}
       \label{eq:root1} \tr_* &=& \frac{2p}{p-4},\\
          \label{eq:root2}     \tr_+ &=& \frac{p}{1+\epsilon},\\
                 \label{eq:root3}\tr_- &=& \frac{p}{1-\epsilon}.
    \end{eqnarray}
\end{subequations}
Note that we avoid naming these roots $\tilde{r}_1$, $\tilde{r}_2$ and $\tilde{r}_3$. The reason is that we reserve numbers to denote that $\tr_1<\tr_2<\tr_3$. On the contrary, the roots in Eq. (\ref{eq:rootsinpe}) can change ordering depending on which branch of $p(\tE,\tL)$ and $\epsilon(\tE,\tL)$ one picks. In other words, if one picks the branch satisfying $p>6+2\epsilon$, then $\tilde{r}_*<\tilde{r}_+<\tilde{r}_-$ and $r_\pm$ are the periapsis and apoapsis radii. However, if we use a different branch for $p$ and $\epsilon$ the roles of each root will be exchanged. 

\begin{figure*}[t]
\centering
  \begin{subfigure}[t]{0.3\textwidth}
    \centering
    \includegraphics[width=\linewidth]{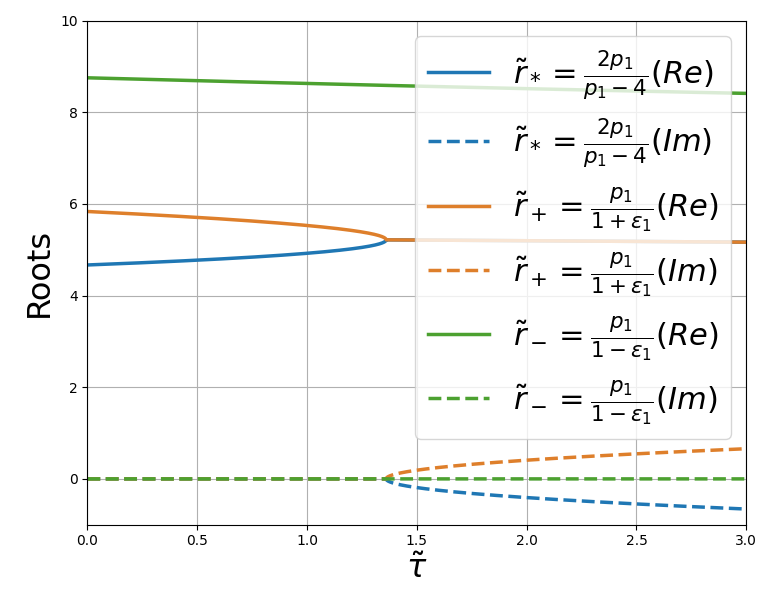}
    \caption{Evolution of roots for first branch of $(p,\epsilon)$.}\label{fig:roots1}
  \end{subfigure}\hfill
  \begin{subfigure}[t]{0.3\textwidth}
    \centering
    \includegraphics[width=\linewidth]{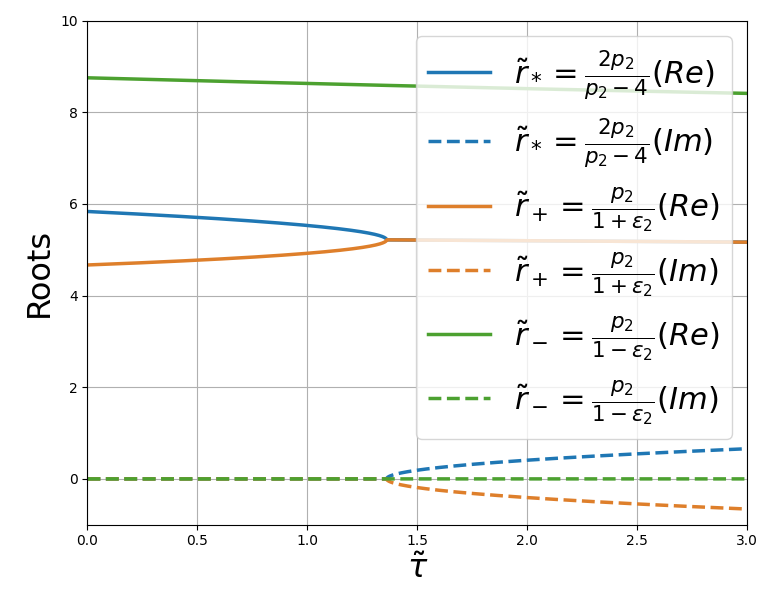}
    \caption{Evolution of roots for second branch of $(p,\epsilon)$.}\label{fig:roots2}
  \end{subfigure}\hfill
  \begin{subfigure}[t]{0.3\textwidth}
    \centering
    \includegraphics[width=\linewidth]{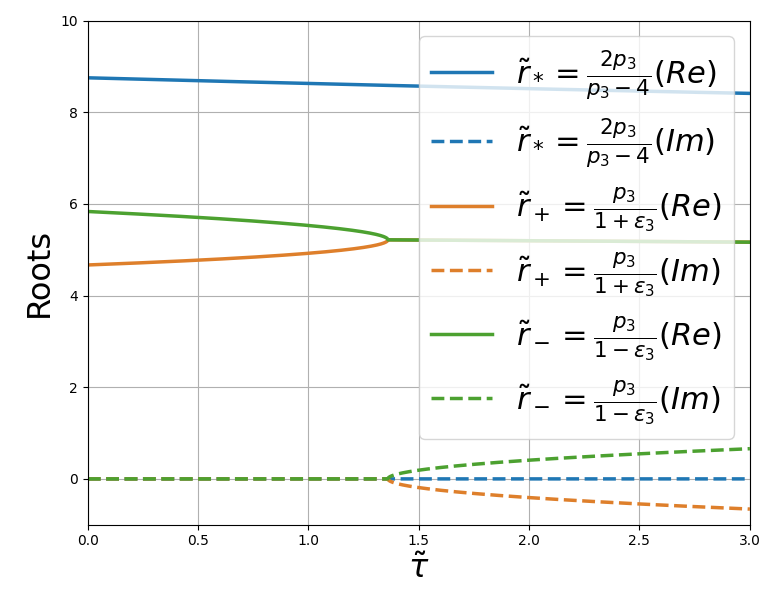}
    \caption{Evolution of roots for third branch of $(p,\epsilon)$.}\label{fig:roots3}
  \end{subfigure}

  \vspace{0.5em}

  \caption{We use simple constant fluxes to evolve $(\tE,\tL)$ across the separatrix. We show the evolution of the three roots $\tr_*$, $\tr_+$ and $\tr_-$ for each branch of $(p,\epsilon)$. Note that, numerically, the three roots always give the same values, although they change identities depending on the branch.}
  \label{fig:rootsevolve}
\end{figure*}

We illustrate this in Fig.~\ref{fig:rootsevolve}, where we evolved $\tE$ and $\tL$ from bound to plunging trajectories using simple constant fluxes. We show the three roots $\tr_*$, $\tr_+$ and $\tr_-$ calculated from the values of $(p,\epsilon)$ on each branch. It is easy to check that the three roots $\tr_*$, $\tr_+$ and $\tr_-$ give, numerically, the same values regardless of what branch is chosen for $p$ and $\epsilon$. However, which root is which changes depending on the branch. We also note that switching between the \emph{first} and \emph{second} branches of $(p,\epsilon)$ has the effect of exchanging $\tr_*$ and $\tr_+$. This will be important in our construction of extended Darwin variables in Section \ref{sec:extendeddarwin}.

\subsection{Limitations of real Darwin variables}

In the usual application of Darwin variables, the first branch for $(p,\epsilon)$ is used. These are restricted to $p>6+2\epsilon$. As the flow of energy and angular momentum drives the system across the separatrix, $p$ and $\epsilon$ become complex, thus producing a complex radius. We propose to solve this issue by analytically continuing the relativistic anomaly $\chi$ in a way that keeps the radius real when $p$ and $\epsilon$ also become complex. In the following section, we show that this approach generalizes the Darwin variables to describe all phases of geodesic motion, not just bound and scattering.

\section{Analytic continuation}\label{sec:analyticcont}

In this section, we extend the domain of validity of Darwin variables by analytically continuing them to the complex plane. We use a subscript to denote real and imaginary parts
\begin{subequations}\label{eq:complexdarwin}
\begin{eqnarray}
    p &=& p_R+ i p_I,\\
    \epsilon &=& \epsilon_R + i \epsilon_I,\\
    \chi &=& \chi_R +i \chi_I.
\end{eqnarray}
\end{subequations}
Replacing the complex variables (\ref{eq:complexdarwin}) in Eq. (\ref{eq:darwinparam}) and Eq. (\ref{eq:eltopepsilon}) will generally produce $\tr$, $\tE^2$ and $\tL^2$ which are complex numbers. Furthermore, we are extending the degrees of freedom from three to six. Hence, we demand that $\tr$, $\tE^2$, and $\tL^2$ remain real and positive to impose three constraints on the extended Darwin variables. The reality conditions on $\tE^2$ and $\tL^2$ can be enforced straightforwardly by parameterizing $(p,\epsilon)$ directly in terms of the physical variables $(\tE,\tL)$; this ensures from the outset that the energy and angular momentum remain real, reducing the number of free degrees of freedom by two. The remaining condition comes from demanding that the radius $\tr$ remains real. We now analyze how this constraints the relativistic anomaly.

\subsection{Reality of $r$}

The radius $r$ in Eq. (\ref{eq:darwinparam}) can be rewritten as
\begin{equation}
    \tr=\frac{p+p\bar{\epsilon} \cos \bar{\chi}}{|1+\epsilon \cos \chi|^2} .
\end{equation}
We now impose that $\Im(\tr)=0$ to get
\begin{equation}\label{eq:Imchiconst}
    \Im \big(\bar{p}\epsilon \cos \chi\big)=p_I ,
\end{equation}
where we applied conjugation to express the final answer in terms of $\chi$ and not $\bar{\chi}$. Eq. (\ref{eq:Imchiconst}) does not determine the real part of $p\bar\epsilon \cos \chi$. We denote the unspecified real part by $\lambda\in \mathbb{R}$. The result is
\begin{equation}\label{eq:chitolambda}
\begin{aligned}
    \cos \chi &= \frac{\lambda+ip_I}{\bar{p}\epsilon}\\
    &= \frac{\alpha}{\alpha^2+\beta^2}\lambda-\frac{\beta\gamma}{\alpha^2+\beta^2}+i\frac{\beta}{\alpha^2+\beta^2}\lambda+\frac{\alpha\gamma}{\alpha^2+\beta^2}.
\end{aligned}
\end{equation}
Here, we define
\begin{subequations}
    \begin{eqnarray}
        \alpha &=& p_R \epsilon_R + p_I \epsilon_I, \\
        \beta &=& p_I\epsilon_R -p_R \epsilon_I,\\
        \gamma &=& p_I.
        \end{eqnarray}
\end{subequations}
First, note that all coefficients are finite. Indeed, they can only diverge if both $|p|^2=|\epsilon|^2=0$, which is not a physical state. Second, Eq. (\ref{eq:chitolambda}) successfully parametrizes a curve in the complex plane in terms of $\lambda$, thus reducing the dimensionality of the complex phase by one. 

Instead of dealing with a complex phase, it is simpler to insert Eq. (\ref{eq:chitolambda}) back into Eq. (\ref{eq:darwinparam}) to get
\begin{equation}\label{eq:freeparam}
    \tr = \frac{|p|^2}{p_R +\lambda}
\end{equation}
which is manifestly real. Note that since $\lambda$ is undetermined, we can simply define
\begin{equation}\label{eq:freeparam2}
    \lambda= \frac{|p|^2}{\tilde \lambda}-p_R
\end{equation}
so that the radius can be parametrized freely by $\tr=\tilde{\lambda}$. This should not be surprising, since the phase $\chi$ in the Darwin variables can be freely reparametrized without affecting the dynamics. Indeed, we are free to reparametrize it as long as it produces radial motion satisfying the correct boundary conditions (i.e. Bound motion between periapsis and apoapsis, etc.).

Eqs. (\ref{eq:freeparam}) and (\ref{eq:freeparam2}) show that there is enough freedom to maintain reality of $(\tr,\tE,\tL)$ by using complex $(p,\epsilon,\chi)$. Motivated by this, we now construct a real parametrization adapted to the different geodesic sectors.

\subsection{Constructing real variables across the separatrix}

In order to choose a sensible parametrization of the radius in terms of $(p,\epsilon)$ it will be convenient to construct quantities that remain real after crossing the separatrix. We will make use of the three roots $\tilde{r}_*$, $\tr_+$ and $\tr_-$ defined in Eq. (\ref{eq:rootsinpe}). We will also make use of switching from the first to the second branch of $(p,\epsilon)$ (cf. Sec. \ref{sec:branches}). Importantly, the first branch of $(p,\epsilon)$ produces $r_*<r_+<r_-$, while the second branch switches the identities of $r_*$ and $r_+$. 

The average between $\tr_*$ and $\tr_+$ can be easily calculated to be
\begin{equation}\label{eq:critradius}
\begin{aligned}
    \tr_{\text{avg}} &= \frac{\tr_*+\tr_+}{2}\\
    &=\frac{p(p+2\epsilon-2)}{2(1+\epsilon)(p-4)}.
    \end{aligned}
\end{equation}
Even though $\tr_{\text{avg}}$ depends on $p$ and $\epsilon$ in a complicated way, it remains real for complex $p$ and $\epsilon$ until the ISCO and is smooth across the separatrix. It loosely follows the position of the UCO at the top of the potential barrier, although it is not exactly it. 

Another quantity of interest is the half-width of the potential barrier at the level of $\tilde{E}^2$
\begin{equation}\label{eq:halfwidth}
\begin{aligned}
    \Delta \tr ^2 &= \left(\frac{\tr_+ -\tr_*}{2}\right)^2\\
    &=\Big[\frac{p(6+2\epsilon-p)}{2(1+\epsilon)(p-4)}\Big]^2.
\end{aligned}
\end{equation}
$\Delta \tr$ is purely real above the separatrix and purely imaginary below. Furthermore, it behaves like the square root of the distance to the separatrix. Hence $\Delta \tr^2$ is always real and behaves linearly at the crossing. 

It is important to note that our expressions (\ref{eq:critradius}) and (\ref{eq:halfwidth}) for $\tr_{\text{avg}}$ and $\Delta \tr$ in terms of $p$ and $\epsilon$ only correspond to the position and half-width of the potential if we restrict to the first two branches of $(p,\epsilon)$. Switching to the second branch only has the effect of switching the sign of $\Delta \tr$, which is why it switches the positions of $\tr_*$ and $\tr_+$. In fact, after crossing the separatrix $\Re(\tr_*)=\Re(\tr_+)=\tr_{\text{avg}}$ and $\Im(\tr_+)=-\Im(\tr_*)=\pm\Delta \tr$.

Although we won't use these, we note that similar constructions can be applied to $p$ and $\epsilon$. In particular, we can define
\begin{subequations}
    \begin{eqnarray}
        p_{\text{avg}}&=&\frac{p_1+p_2}{2},\\
        \Delta p &=&\frac{p_1-p_2}{2},
    \end{eqnarray}
\end{subequations}
where $p_1$ and $p_2$ are the first and second branches depicted in Fig. \ref{fig:pbranches}. Then, $p_{\text{avg}}$ and $\Delta p^2$ are real across the separatrix. An analogous construction works for the eccentricity as well. 

\section{Extended Darwin variables across bound, scattering, and plunging sectors}\label{sec:extendeddarwin}

We now construct an explicit real parametrization that reproduces the correct turning-point structure in all sectors of motion.

We parameterize the radius as
\begin{equation}\label{eq:radiusparamallcases}
    \tr = \frac{1}{f(p,\epsilon)+A(p,\epsilon)\varphi(\eta)},
\end{equation}
where $f$ and $A$ are
\begin{subequations}\label{eq:fAdef}
    \begin{eqnarray}
        f &=& \frac{1}{2}\left(\frac{1}{\Re(\tr_+)}+\frac{1}{\Re(\tr_-)}\right),\label{boundf}\\
        A &=& \frac{1}{2}\left(\frac{1}{{\Re(\tr_+)}}-\frac{1}{\Re(\tr_-)}\right).\label{boundA}
        \end{eqnarray}
\end{subequations}
The function $\varphi(\eta)$ will change depending on the type of motion
\begin{equation}\label{eq:varphidef}
    \varphi(\eta)=\begin{cases} \cos\big(\eta\big)& \text{for bound/scattering trajectories},\\
    \cosh\big(\eta-\eta_d\big)-2 & \text{for outer/direct plunges},\\
    \cosh\big(\eta-\eta_t\big) & \text{for inner plunges}.
    \end{cases}
\end{equation}
The initial phases $\eta_d$ and $\eta_t$ can be freely specified if one is parametrizing each phase separately. However, if one includes dissipation and the system evolves from one phase to the next, the initial phases need to be chosen carefully as to make $\varphi(\eta)$ smooth across the separatrix.

The parametrization in Eq. (\ref{eq:radiusparamallcases}) is not uniquely determined by the analytic continuation, but is chosen so as to reproduce the correct turning points for bound, scattering and plunging trajectories.

\subsection{Bound and scattering trajectories}\label{subsec:bound}

These trajectories correspond to values of $(\tE,\tL)$ above the separatrix. We use the first branch of $(p,\epsilon)$. On this phase, all roots are real. Then, simple algebra reduces our expression to $f=1/p$ and $A=\epsilon/p$ and we recover the usual Darwin variables.

In this choice of coordinates, the particle passes through periapsis at $\eta=2\pi n$ and apoapsis at $\eta=(2n+1)\pi$.

\subsection{Outer and direct plunges}\label{subsec:direct}

These trajectories correspond to values of $(\tE,\tL)$ below the separatrix. We will use the first branch of $(p,\epsilon)$, for which $\tr_*$ and $\tr_+$ are complex but $\tr_-$ remains real. Taking the real part of $\tr_+$ returns the position of the potential barrier $\tr_{\text{avg}}$. 

With the parametrization (\ref{eq:radiusparamallcases}), we get that the radius is $\tr=\tr_-$ at $\eta=\eta_0$ and then goes to $\tr=0$ for either $\eta\rightarrow \pm \infty$.

In the case of direct plunges, $\tr_-$ is negative. This means that the physical trajectory won't start on $\eta_0$, but on a later time at which $\tr\rightarrow\infty$. This is the same behaviour for scattering trajectories in the usual Darwin variables, where the negative root represents the ``turning point at infinity''. 

\subsection{Inner plunges}\label{subsec:trapped}

These trajectories correspond to values of $(\tE,\tL)$ above the separatrix but radii to the left of the potential barrier. All roots are real. We use the second branch of $(p,\epsilon)$ which flips the identities of $\tr_*$ and $\tr_+$ so that $\tr_+$ becomes the smallest root and $\tr_*$ the periapsis. 

Then, we obtain that the radius starts at $\tr_+$ at $\eta=\eta_0$ and then goes to zero for either $\eta\rightarrow \pm \infty$. We note the importance of changing branches to make this work. Otherwise, the particle would start at the periapsis and cross the forbidden region where energy $\tE$ is lower than the potential barrier.

\subsection{Plunges below the ISCO}

In the case of quasi-circular inspirals, the system can cross the separatrix exactly at the ISCO. After the ISCO is crossed, the effective potential no longer exhibits a barrier and the radial motion requires a careful treatment. For energies $\tE<1$, the radial equation admits one real root and two complex-conjugate roots. Which one of the three roots remains real (and thus represents the turning point) depends on whether the turning point is smaller or greater than $\tr_{\text{ISCO}}=6$. This, in turn, depends on the relative strength of the energy and angular-momentum fluxes. If the energy flux dominates, the turning point moves inward toward the central black hole. In this situation, the smallest root $\tr_1$ remains real and defines the turning point. If instead the angular-momentum flux dominates, the shape of the effective potential is modified so that the turning point moves outward. In that case, the largest root $\tr_3$ remains real and becomes the turning point.

Hence, when the turning point moves outwards, the trajectory behaves as a \emph{outer plunge} (cf. subsection \ref{subsec:direct}). Instead, when the turning point moves inwards, the motion corresponds to a \emph{inner plunge} (cf. subsection \ref{subsec:trapped}). Therefore, the plunges occurring after the ISCO has been crossed can also be described by the parametrization given by Eq. (\ref{eq:fAdef}).

\section{Evolution under simple driving force}\label{sec:rad}

We are interested in applying the extended Darwin variables for systems where a small driving force evolves the system from one phase to another. Specifically, we want the system to start on a bound trajectory and slowly lose energy and angular momentum until it transitions to either an inner or outer plunge. The transition occurs when the difference between $\tE^2$ and the maximum of the potential barrier is zero. We follow the work in \cite{Faggioli.xq3j-4jtx} and define this difference as 
\begin{equation}
\Delta=\tE^2-V_{\text{eff}}(\tr_{\text{UCO}};\tL).
\end{equation}
Hence, $\Delta$ is negative for bound orbits, scattering trajectories and inner plunges and positive for outer and direct plunges. We note that a bound orbit can connect continuously to an inner plunge only in the fine-tuned case where the separatrix is crossed while the particle is at periapsis. Otherwise, inner plunges remain disconnected from bound and scattering trajectories. 

Now, let us consider a transition to an outer plunge. Given a separatrix crossing phase $\eta_{\text{sep}}$, we define the closest passage through apoapsis as
\begin{subequations}
\begin{eqnarray}
     n &=& \left \lfloor\frac{\eta_\text{sep}}{2\pi} \right \rfloor,\\
    \eta_{\text{apo}} &=& (2n+1)\pi.
\end{eqnarray}
\end{subequations}
For example, both $\eta_\text{sep}=0.1 \pi$ and $\eta_\text{sep}=1.1 \pi$ give $\eta_{\text{apo}}=\pi$ .
Then, $\varphi(\eta)$ is
\begin{equation}\label{eq:transtodirect}
    \varphi(\eta)=\begin{cases} \cos(\eta) & \eta<\eta_{\text{apo}},\\
    \cosh(\eta-\eta_{\text{apo}})-2 & \eta>\eta_{\text{apo}}. 
    \end{cases}
\end{equation}
This parametrization is continuous up to its second derivative.

Note that we can switch parametrizations either before or after the actual crossing of the separatrix. This will depend on whether the particle had positive or negative radial velocity at the time of crossing. When the particle has positive radial velocity, we are extending the use of the parametrization for bound  orbits past the crossing of the separatrix. This is not a problem since both bound orbits and outer plunges have a turning point at $\tr_3$.

Consider now the transition to an inner plunge. In this case, the crossing of the separatrix must happen at a time $\eta_{\text{sep}}=2\pi n$. We simply define
\begin{equation}\label{eq:transtotrapped}
    \varphi(\eta)=\begin{cases} \cos(\eta) & \eta<\eta_{\text{sep}},\\
    \cosh(\eta-\eta_{\text{sep}}) & \eta>\eta_\text{sep}.
    \end{cases}
\end{equation}

We note that this parametrization has a discontinuity on its first derivative. This is to be expected as the particle is crossing a critical point that limits bound and plunging trajectories. This can be traced back to the fact that the particle would spend an infinite amount of proper time at the critical point, unless perturbed by an external force. In fact, there is a more serious problem with this transition, as we discuss in the next subsection.

Finally, we also note that it is possible, in theory, for a system to start with $\Delta<0$ on a bound orbit, then cross the separatrix into an outer plunge with $\Delta >0$ and then transition back into an inner plunge with $\Delta<0$ after crossing the UCO. This can occur if the angular momentum flux exceeds the energy flux in such a way that the gap $\Delta$ closes back. Work in \cite{Faggioli.xq3j-4jtx} observed this type of behavior in numerical simulations, but the presence of this effect in astrophysical systems remains unclear. The coordinates defined in Eqs. (\ref{eq:fAdef}) and (\ref{eq:varphidef}) cannot be made easily continuous in a transition from outer to inner plunges. However, this is not necessary. After the UCO has been crossed, outer and inner plunges behave in the same way: The radius decreases monotonically down to the horizon. Hence, even if the system switches between the two phases, there is no need to change the parametrization.

\subsection{Kink at the separatrix crossing}

Even though our parametrization is continuous across the separatrix, it depends on $\Re(\tr_+)$ which behaves like $\sqrt{|\eta_{\text{sep}}-\eta}|$, as  can be seen in Fig. \ref{fig:roots1}. Hence, its first time derivative is infinite at the crossing. This is a problem when attempting to do an adiabatic transition from bound/scattering to outer/direct plunges, as the change in $\tr_+$ will dominate over the slow changes in $\tE$ and $\tL$. 

\begin{figure}[h]
    \centering
\includegraphics[width=1\linewidth]{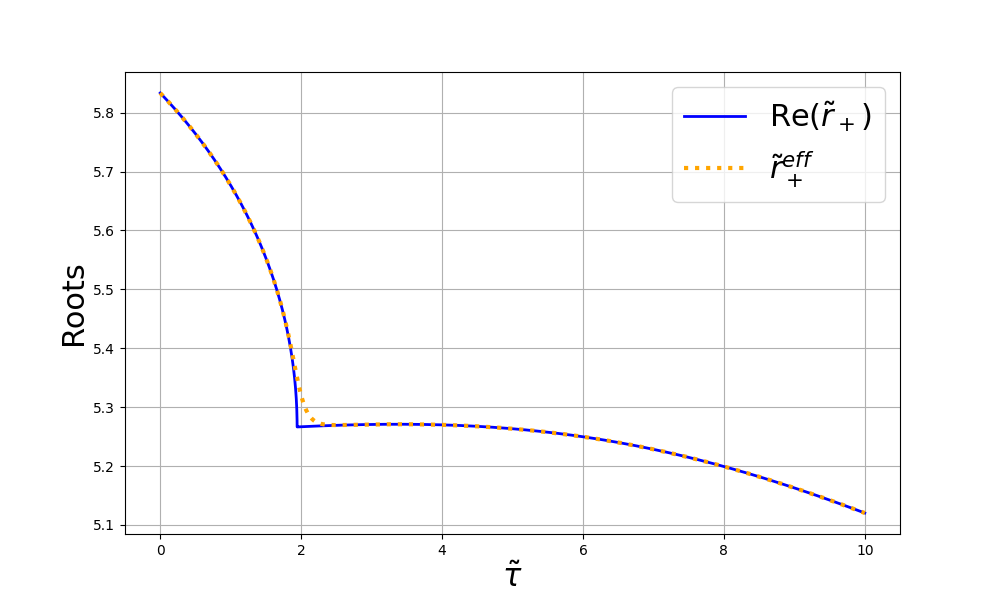}
    \caption{Comparison between the root $\tr_+$ and the smoothed root $\tr_+^{\text{eff}}$ defined in Eqs. (\ref{eq:effrootdef}) and (\ref{eq:effrootdef2}). We choose $l=1/100$ as the length cutoff.}
    \label{fig:effectiveroot}
\end{figure}

However, note that the parametrization for outer/direct plunges continues to have the correct turning point if one drops the first term in $f$ and $A$ in Eq. (\ref{eq:fAdef}). Indeed, that term is only there for continuity with the bound/scattering trajectories. Hence, it is possible to fix this for most realistic situations by replacing $\tr_+$ in Eq. (\ref{eq:fAdef}) with a smoothed variable
\begin{equation}\label{eq:effrootdef}
    \tr_+^{\text{eff}}= \tr_{\text{avg}} +\sigma(\Delta \tr^2) .
\end{equation}
Here, the function $\sigma(\Delta \tr^2)$ should interpolate between $\Delta\tr$ and zero in a smooth way. We propose
\begin{equation}\label{eq:effrootdef2}
\sigma_l(x)=l\sqrt{\ln\big[1+e^{x/l^2}\big]}.
\end{equation}
This expression satisfies $\sigma_l(x)\approx \sqrt{x}$ for $x\gg l^2$, $\sigma_l(0)\propto l $ and $\sigma_l(x) \approx l e^{x/2l^2}\rightarrow 0$ for $x\ll-l^2$. We show both $\tr_+$ and $\tr_+^{\text{eff}}$ on Fig. \ref{fig:effectiveroot}.

For most values of $\Delta \tr$, the effective root $\tr_+^{\text{eff}}$ will be equal to the actual root $\tr_+$ up to exponentially small corrections. As $\Delta \tr$ approaches zero, the effective variable approaches $\tr_{\text{avg}}$ smoothly, instead of through a square root kink. Whenever the transition doesn't occur too close to periapsis, this doesn't affect the dynamics. However, the dynamics become sensitive to the arbitrary choice of cutoff $l$ if the separatrix is crossed at a radius $\tr$ such that the distance $\tr-\tr_{\text{UCO}}\approx O(l)$.

We note that this approach doesn't work for transitions to inner plunges, since these occur precisely at the periapsis. Hence, the parametrization (\ref{eq:transtotrapped}) is flawed. We leave an understanding of how to choose smooth variables for this transition for future work. However, these transitions are fine tuned; and generic transitions will always occur at some finite distance from periapsis, which makes this approach useful for most physically relevant situations.

\subsection{Toy model for adiabatic evolution under driving force}

In order to test how the coordinates evolve across the separatrix, we evolve $\tE$ and $\tL$ at a constant rate to drive a point particle from a bound orbit into an outer plunge. At each time, we make the particle follow a geodesic whose parameters $(p,\epsilon)$ are determined by the slowly varying $(\tE,\tL)$. We parametrize the radius using Eqs. (\ref{eq:radiusparamallcases}) and (\ref{eq:transtodirect}).

\begin{figure*}[t]
\centering

\begin{subfigure}[t]{0.45\textwidth}
    \centering
    \includegraphics[width=\linewidth]{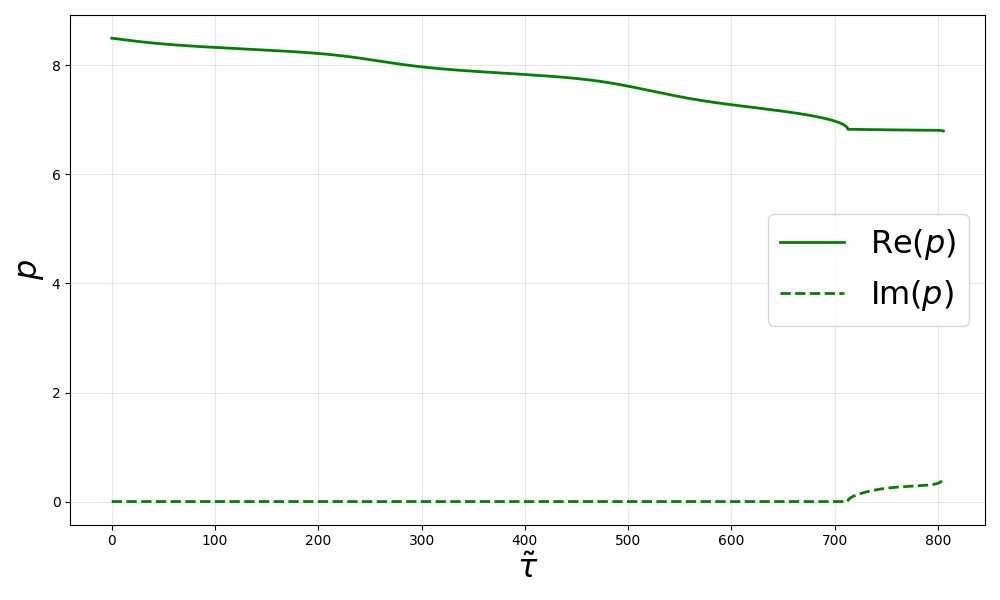}
    \caption{Evolution of semi-latus rectum $p$.}
    \label{fig:toyp}
\end{subfigure}
\hfill
\begin{subfigure}[t]{0.45\textwidth}
    \centering
    \includegraphics[width=\linewidth]{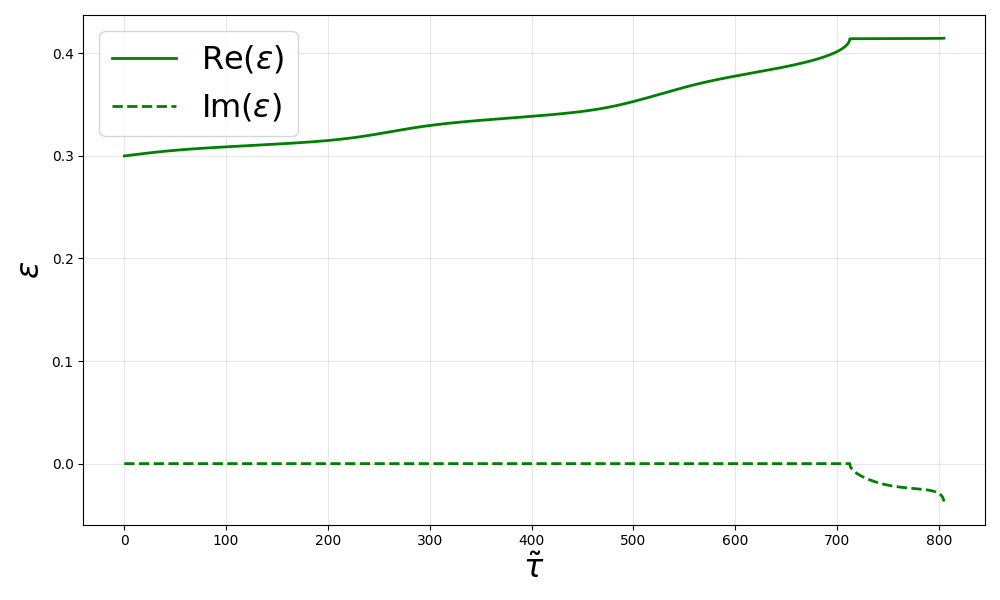}
    \caption{Evolution of eccentricity $\epsilon$.}
    \label{fig:toyecc}
\end{subfigure}

\vspace{0.5em}

\begin{subfigure}[t]{0.45\textwidth}
    \centering
    \includegraphics[width=\linewidth]{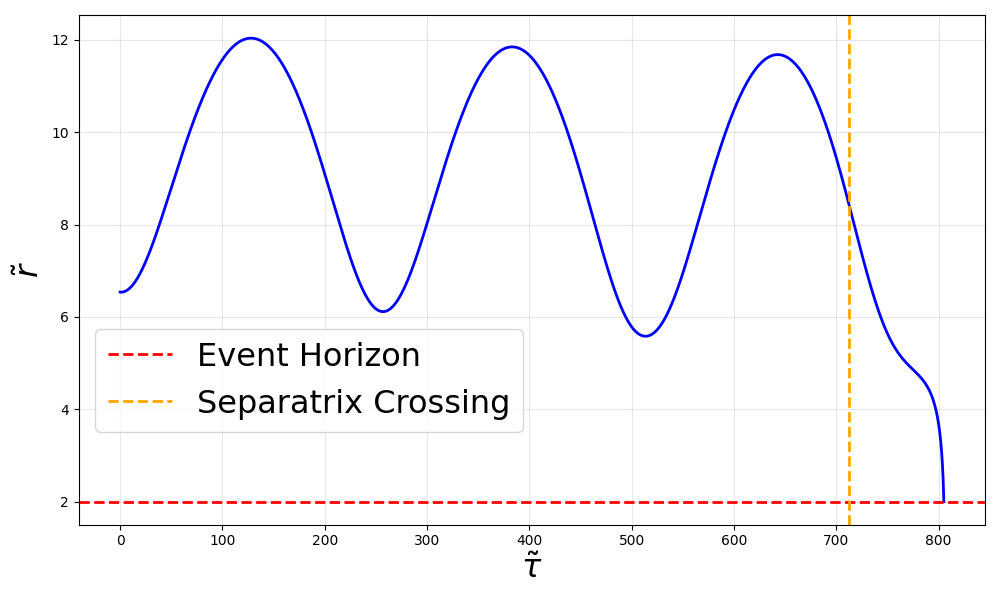}
    \caption{Evolution of radius.}
    \label{fig:toyr}
\end{subfigure}
\hfill
\begin{subfigure}[t]{0.45\textwidth}
    \centering
    \includegraphics[width=\linewidth]{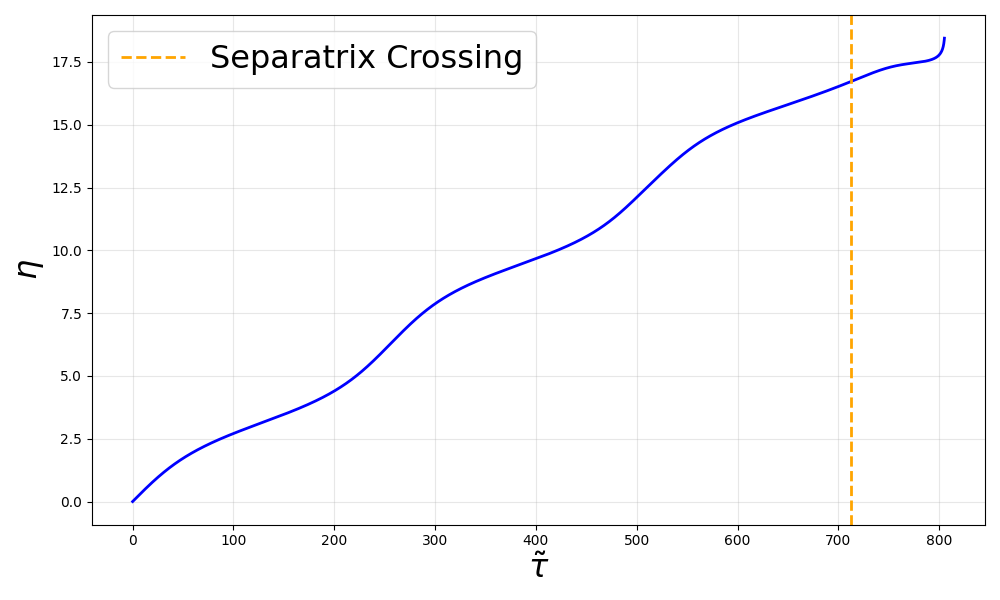}
    \caption{Evolution of phase $\eta$ as a function of proper time $\tilde{\tau}$.}
    \label{fig:toyeta}
\end{subfigure}

\caption{We use simple constant fluxes to evolve $(\tilde E,\tilde L)$ and make a particle follow, at each time, a geodesic with values $p(\tE,\tL)$ and $\epsilon(\tE,\tL)$. Even though $(p,\epsilon)$ have a kink at the separatrix and become complex after that, the radius $\tr$ and orbital phase $\eta$ remain real and smooth throughout the complete evolution of the dynamics.}
\label{fig:toymodel}
\end{figure*}

We chose constant rates of change $d\tilde{E}/d\eta=-1.5\times10^{-4}$ and $d\tL/d\eta = -5\times10^{-3}$. These rates were chosen only to illustrate a transition to an outer plunge; they are not intended to model a realistic physical system. We start the particle with semi-latus rectum $p_0=8.5$ and eccentricity $\epsilon=0.3$. In Fig. \ref{fig:toymodel}, we show $p$ and $\epsilon$ as functions of proper time, as well as the radius $\tr$ and the phase $\eta$. This demonstrates that the extended variables provide a continuous parametrization of the motion across the separatrix, even though the underlying orbital elements $(p,\epsilon)$ develop non-analytic behavior.

\section{Conclusions}

In this work, we analytically continued the Darwin variables to include plunging geodesics. For geodesic motion, the extended variables provide a continuous description of bound, scattering and plunge trajectories. In the presence of a driving force, a smooth parametrization can be maintained except in fine-tuned transitions occurring arbitrarily close to periapsis. We leave the use of these new coordinates on dynamical systems with realistic radiation-reaction forces for future work. We also note that it should be possible to extend these coordinates to geodesics on a Kerr background.

\vspace{0.1cm}
\noindent

\textit{Acknowledgments}: We thank Maarten Van de Meent, Guillaume Lhost, Guglielmo Faggioli and Devin Becker for valuable discussions and insights into the lore of the transition to plunge. We also thank Nami Nishimura and Trevor Scheopner for fruitful discussions on the analytic continuation on Darwin variables. \emph{Funded by the European Union. Views and opinions expressed are however those of the author(s) only and do not necessarily reflect those of the European Union or the European Research Council Executive Agency. Neither the European Union nor the granting authority can be held responsible for them. This work is supported by ERC grant (GWSky/ 101167314).}


\twocolumngrid

\newpage
\bibliography{Ref.bib}

@article{Damour:2000we,
title = {Dimensional regularization of the gravitational interaction of point masses},
journal = {Physics Letters B},
volume = {513},
number = {1},
pages = {147-155},
year = {2001},
issn = {0370-2693},
doi = {https://doi.org/10.1016/S0370-2693(01)00642-6},
url = {https://www.sciencedirect.com/science/article/pii/S0370269301006426},
author = {Thibault Damour and Piotr Jaranowski and Gerhard Schäfer}
}

@article{Buonanno:2025xu,
  title = {Transition from inspiral to plunge in precessing binaries of spinning black holes},
  author = {Buonanno, Alessandra and Chen, Yanbei and Damour, Thibault},
  journal = {Phys. Rev. D},
  volume = {74},
  issue = {10},
  pages = {104005},
  numpages = {26},
  year = {2006},
  month = {Nov},
  publisher = {American Physical Society},
  doi = {10.1103/PhysRevD.74.104005},
  url = {https://link.aps.org/doi/10.1103/PhysRevD.74.104005}
}

@article{darwin,
 ISSN = {00804630},
 URL = {http://www.jstor.org/stable/100508},
 author = {Charles Darwin},
 journal = {Proceedings of the Royal Society of London. Series A, Mathematical and Physical Sciences},
 number = {1257},
 pages = {180--194},
 publisher = {The Royal Society},
 title = {The Gravity Field of a Particle},
 urldate = {2026-03-27},
 volume = {249},
 year = {1959}
}

@article{Levin:2008yp,
    author = "Levin, Janna and Perez-Giz, Gabe",
    title = "{Homoclinic Orbits around Spinning Black Holes. I. Exact Solution for the Kerr Separatrix}",
    eprint = "0811.3814",
    archivePrefix = "arXiv",
    primaryClass = "gr-qc",
    doi = "10.1103/PhysRevD.79.124013",
    journal = "Phys. Rev. D",
    volume = "79",
    pages = "124013",
    year = "2009"
}

@article{Grossman:2008yk,
    author = "Grossman, Rebecca and Levin, Janna",
    title = "{Dynamics of Black Hole Pairs II: Spherical Orbits and the Homoclinic Limit of Zoom-Whirliness}",
    eprint = "0811.3798",
    archivePrefix = "arXiv",
    primaryClass = "gr-qc",
    doi = "10.1103/PhysRevD.79.043017",
    journal = "Phys. Rev. D",
    volume = "79",
    pages = "043017",
    year = "2009"
}

@article{Buonanno_1999,
  title = {Effective one-body approach to general relativistic two-body dynamics},
  author = {Buonanno, A. and Damour, T.},
  journal = {Phys. Rev. D},
  volume = {59},
  issue = {8},
  pages = {084006},
  numpages = {24},
  year = {1999},
  month = {Mar},
  publisher = {American Physical Society},
  doi = {10.1103/PhysRevD.59.084006},
  url = {https://link.aps.org/doi/10.1103/PhysRevD.59.084006}
}

@article{Lhost:2024jmw,
    author = "Lhost, Guillaume and Comp{\`e}re, Geoffrey",
    title = "{Approach to the separatrix with eccentric orbits}",
    eprint = "2412.04249",
    archivePrefix = "arXiv",
    primaryClass = "gr-qc",
    doi = "10.21468/SciPostPhysCore.8.3.059",
    journal = "SciPost Phys. Core",
    volume = "8",
    pages = "059",
    year = "2025"
}

@article{Becker.PhysRevD.111.064003,
  title = {Transition from adiabatic inspiral to plunge for eccentric binaries},
  author = {Becker, Devin R. and Hughes, Scott A.},
  journal = {Phys. Rev. D},
  volume = {111},
  issue = {6},
  pages = {064003},
  numpages = {22},
  year = {2025},
  month = {Mar},
  publisher = {American Physical Society},
  doi = {10.1103/PhysRevD.111.064003},
  url = {https://link.aps.org/doi/10.1103/PhysRevD.111.064003}
}

@article{Faggioli.xq3j-4jtx,
  title = {Characterizing the merger of equatorial-eccentric-geodesic plunges in rotating black holes},
  author = {Faggioli, Guglielmo and van de Meent, Maarten and Buonanno, Alessandra and Khanna, Gaurav},
  journal = {Phys. Rev. D},
  volume = {112},
  issue = {8},
  pages = {084009},
  numpages = {29},
  year = {2025},
  month = {Oct},
  publisher = {American Physical Society},
  doi = {10.1103/xq3j-4jtx},
  url = {https://link.aps.org/doi/10.1103/xq3j-4jtx}
}

@article{Kuchler:2024esj,
    author = {K{\"u}chler, Lorenzo and Comp{\`e}re, Geoffrey and Durkan, Leanne and Pound, Adam},
    title = "{Self-force framework for transition-to-plunge waveforms}",
    eprint = "2405.00170",
    archivePrefix = "arXiv",
    primaryClass = "gr-qc",
    doi = "10.21468/SciPostPhys.17.2.056",
    journal = "SciPost Phys.",
    volume = "17",
    number = "2",
    pages = "056",
    year = "2024"
}

@article{Honet:2025dho,
    author = {Honet, Lo{\"\i}c and K{\"u}chler, Lorenzo and Pound, Adam and Comp{\`e}re, Geoffrey},
    title = "{Transition-to-plunge self-force waveforms with a spinning primary}",
    eprint = "2510.13958",
    archivePrefix = "arXiv",
    primaryClass = "gr-qc",
    doi = "10.1103/sq6y-qv8h",
    journal = "Phys. Rev. D",
    volume = "113",
    number = "4",
    pages = "044051",
    year = "2026"
}

@article{Küchler_2026,
doi = {10.1088/1361-6382/ae2b44},
url = {https://doi.org/10.1088/1361-6382/ae2b44},
year = {2025},
month = {dec},
publisher = {IOP Publishing},
volume = {43},
number = {1},
pages = {015018},
author = {Küchler, Lorenzo and Compère, Geoffrey and Pound, Adam},
title = {Self-force framework for merger-ringdown waveforms},
journal = {Classical and Quantum Gravity}
}

@article{PhysRevD.62.124022,
  title = {Transition from inspiral to plunge for a compact body in a circular equatorial orbit around a massive, spinning black hole},
  author = {Ori, Amos and Thorne, Kip S.},
  journal = {Phys. Rev. D},
  volume = {62},
  issue = {12},
  pages = {124022},
  numpages = {8},
  year = {2000},
  month = {Nov},
  publisher = {American Physical Society},
  doi = {10.1103/PhysRevD.62.124022},
  url = {https://link.aps.org/doi/10.1103/PhysRevD.62.124022}
}

@article{PhysRevD.62.064015,
  title = {Transition from inspiral to plunge in binary black hole coalescences},
  author = {Buonanno, Alessandra and Damour, Thibault},
  journal = {Phys. Rev. D},
  volume = {62},
  issue = {6},
  pages = {064015},
  numpages = {24},
  year = {2000},
  month = {Aug},
  publisher = {American Physical Society},
  doi = {10.1103/PhysRevD.62.064015},
  url = {https://link.aps.org/doi/10.1103/PhysRevD.62.064015}
}

@article{introLIGO1,
    author = "Abbott, B. P. and others",
    collaboration = "LIGO Scientific, Virgo",
    title = "{Observation of Gravitational Waves from a Binary Black Hole Merger}",
    eprint = "1602.03837",
    archivePrefix = "arXiv",
    primaryClass = "gr-qc",
    reportNumber = "LIGO-P150914",
    doi = "10.1103/PhysRevLett.116.061102",
    journal = "Phys. Rev. Lett.",
    volume = "116",
    number = "6",
    pages = "061102",
    year = "2016"
}

@article{introLIGO2,
    author = "Abbott, B. P. and others",
    collaboration = "LIGO Scientific, Virgo",
    title = "{GW151226: Observation of Gravitational Waves from a 22-Solar-Mass Binary Black Hole Coalescence}",
    eprint = "1606.04855",
    archivePrefix = "arXiv",
    primaryClass = "gr-qc",
    reportNumber = "LIGO-P151226",
    doi = "10.1103/PhysRevLett.116.241103",
    journal = "Phys. Rev. Lett.",
    volume = "116",
    number = "24",
    pages = "241103",
    year = "2016"
}

@article{introLIGO3,
    author = "Abbott, B. P. and others",
    collaboration = "LIGO Scientific, Virgo",
    title = "{Binary Black Hole Mergers in the first Advanced LIGO Observing Run}",
    eprint = "1606.04856",
    archivePrefix = "arXiv",
    primaryClass = "gr-qc",
    reportNumber = "LIGO-P1600088",
    doi = "10.1103/PhysRevX.6.041015",
    journal = "Phys. Rev. X",
    volume = "6",
    number = "4",
    pages = "041015",
    year = "2016"
}

@article{introLISA1,
    author = "{P. Amaro-Seoane et al.}",
    title = "{}",
    eprint = "1305.5720",
    archivePrefix = "arXiv",
    primaryClass = "astro-ph.CO",
    doi = "",
    journal = "",
    volume = "",
    number = "",
    pages = "",
    year = "2013"
}

@article{flanagan2,
    author = "Hinderer, Tanja and Flanagan, Eanna E.",
    title = "{Two timescale analysis of extreme mass ratio inspirals in Kerr. I. Orbital Motion}",
    eprint = "0805.3337",
    archivePrefix = "arXiv",
    primaryClass = "gr-qc",
    doi = "10.1103/PhysRevD.78.064028",
    journal = "Phys. Rev. D",
    volume = "78",
    pages = "064028",
    year = "2008"
}

@article{pound,
    author = "Barack, Leor and Pound, Adam",
    title = "{Self-force and radiation reaction in general relativity}",
    eprint = "1805.10385",
    archivePrefix = "arXiv",
    primaryClass = "gr-qc",
    doi = "10.1088/1361-6633/aae552",
    journal = "Rept. Prog. Phys.",
    volume = "82",
    number = "1",
    pages = "016904",
    year = "2019"
}

\end{document}